# Evaluating the Predictive Capacity of ChatGPT for Academic Peer Review Outcomes Across Multiple Platforms

Mike Thelwall, Abdallah Yaghi: Information School, University of Sheffield, UK.

Abstract

Academic peer review is at the heart of scientific quality control, yet the process remains slow, time-consuming, and increasingly strained by rising academic workloads. Technology that can predict peer review outcomes may help with this, for example by fast-tracking desk rejection decisions. While previous studies have demonstrated that Large Language Models (LLMs) can predict peer review outcomes to some extent, this paper builds on that by introducing two new contexts and employing a more robust method—averaging multiple ChatGPT scores. The findings that averaging 30 ChatGPT predictions, based on reviewer guidelines and using only the submitted titles and abstracts, failed to predict peer review outcomes for F1000Research (Spearman's rho = 0.00). However, it produced mostly weak positive correlations with the quality dimensions of SciPost Physics (rho = 0.25 for validity, rho = 0.25 for originality, rho = 0.20 for significance, and rho = 0.08 for clarity) and a moderate positive correlation for papers from the International Conference on Learning Representations (ICLR) (rho = 0.38). Including the full text of articles significantly increased the correlation for ICLR (rho = 0.46) and slightly improved it for F1000Research (rho = 0.09), while it had variable effects on the four quality dimension correlations for SciPost LaTeX files. The use of chain-of-thought system prompts slightly increased the correlation for F1000Research (rho = 0.10), marginally reduced it for ICLR (rho = 0.37), and further decreased it for SciPost Physics (rho = 0.16 for validity, rho = 0.18 for originality, rho = 0.18 for significance, and rho = 0.05 for clarity). Overall, the results suggest that in some contexts, ChatGPT can produce weak pre-publication quality assessments. However, the effectiveness of these assessments and the optimal strategies for employing them vary considerably across different platforms, journals, and conferences. Additionally, the most suitable inputs for ChatGPT appear to differ depending on the platform.

## Introduction

In theory, academic research undergoes peer review before being validated through publication in a recognised journal, full-text conference, monograph, edited book, or alternative publishing platform. While this quality control mechanism is essential, it remains imperfect (Hemlin, 2009) and faces competition from the preprint publication model (Wingen et al., 2022). At the very least, peer review serves to identify work that demonstrates higher interest, originality, and/or validity compared to other research outputs. However, this process depends on active scholars reviewing each other's work, a task that is typically time-consuming and conducted without financial compensation (Aczel et al., 2021). With the growing volume of academic publications, as recorded by leading international bibliometric databases (Bornmann et al., 2021), and the apparent increase in managerial control over academic time (e.g., McCarthy, & Dragouni, 2021), the sustainability of the current system is under threat. This is evidenced by growing difficulties in securing reviewers in certain fields (personal experience; see also: Franceschet et al., 2022; Mogaji, 2024), even though AI is often used to help find potential reviewers (Checco et al., 2021). In this context, strategies to alleviate the workload of peer reviewers would be highly advantageous, with Artificial Intelligence (AI) technology offering a promising solution. For instance, the La Caixa Foundation employs AI to identify grant applications that are unlikely to be successful, with these decisions subsequently reviewed by two human experts before a desk rejection is confirmed (Carbonell Cortés, 2024). This approach not only supports editors in reducing reviewer workloads but also maintains ethical standards, provided that authors are informed and that the applications are not publicly disclosed or used to train Large Language Models (LLMs).

Previous research has also shown that AI can offer valuable assistance to reviewers by providing suggestions regarding the content of their reviews (Liang et al., 2024b), identifying reviews generated by LLM (Liang et al., 2024a) or rewarding reviewers for producing high-quality reviews (Lu et al., 2024). AI has been shown to provide suggestions comparable to those of human reviewers (Liang et al., 2024b), meaning that editors and reviewers could potentially benefit from using AI as a tool to ensure they have not overlooked key elements in their initial review. However, relying on AI as the primary source for a review remains problematic, particularly because the responsibility for ensuring the accuracy of a review must rest with human agent. The use of human reviewers mitigates the risk of publishing articles that are clearly flawed or deceptive to human readers (although human reviewers also make mistakes: Bar-Ilan & Halevi, 2018). Additionally, human reviewers also act as a safeguard against authors who may attempt to manipulate AI, such as by requesting the AI to rewrite their paper in a way that would result in a more favourable evaluation. AI could, however, be effectively employed for meta-reviewing, in the sense of synthesising or evaluating the reviews provided by human reviewers (Du et al., 2024) and generating summaries to improve efficiency in the peer review process (Chauhan & Currie, 2024).

LLMs have also shown some ability to predict peer review outcomes in some contexts, often providing explanation to support their decision outcome. For example, ChatGPT scores ranging from 1 to 5 have been shown to weakly correlate with reviewer scores for papers submitted to the International Conference on Learning Representations (ICLR) 2017, based on titles and abstracts (Spearman: 0.282). However, when full texts were entered, the system appeared to become confused, resulting in a weaker correlation (Spearman: 0.091) (Zhou et al., 2024), despite the necessity of full-text access for comprehensive peer review. Nonetheless, this context is narrow, and the broader relevance of AI prediction capabilities remains unclear.

From a different perspective, previous research on post-publication expert review quality scoring by ChatGPT (n=51 articles) has demonstrated that individual tests (Spearman correlation: 0.38) yield lower correlations with human evaluations compared to averaging 30 ChatGPT predictions. The research further confirms that providing full texts (Spearman: 0.60) results in lower correlations than using only titles and abstracts (Spearman: 0.67) (Thelwall, 2024ab). It is thus reasonable to conclude that averaging predictions may also enhance accuracy in pre-publication peer review assessments.

Finally, numerous studies have analysed prompting strategies for LLMs. In the context of academic peer review, a prompt might be as simple as describing the scoring system (Zhou et al., 2024). Nevertheless, it seems more appropriate to provide the AI with the same instructions given to human reviewers (Thelwall, 2024ab), as this would succinctly describe the task. That said, for more complex tasks, Chain-of-Thought prompting has been shown to be particularly effective for LLMs, suggesting a potential different approach.

Chain-of-thought prompting involves creating system prompts that guide the model through a step-by-step process, breaking down a multi-stage problem in a manner that allows the system to assemble the necessary information to make more informed decisions (Zhang et al., 2022), although it seems to work mainly on early version of ChatGPT or other LLMs. It is not clear how well this applies to ChatGPT 3.5+ versions however, since they are optimised for following natural language instructions for tasks. Nevertheless, it may be worth investigating whether further accuracy could be achieved by restructuring the system prompts to request the decision at the end of the report. This mirrors the chain-of-thought approach by encouraging the system to generate relevant information before reaching its final decision.

Based on the above discussion, this paper addresses the following questions. The focus is on ChatGPT, as there is currently no evidence to suggest that competing systems offer superior performance in peer review tasks.

- RQ1: Does ChatGPT have some ability to predict pre-publication peer review decisions across various contexts?
- RQ3: Does averaging ChatGPT predictions consistently yield more accurate results for pre-publication peer review than individual predictions?
- RQ3: Are system prompts more effective if they ask for scores to be suggested at the end of reports?

## Methods

The research design involved three key steps: (1) downloading large sets of pre-publication academic documents, along with their corresponding reviewer recommendations, from various contexts, (2) requesting ChatGPT predictions using its Application Programming Interface (API), and (3) correlating the averaged ChatGPT predictions with reviewer outcomes to address the research questions.

### Data

Although peer review reports and outcomes are typically confidential, they are occasionally made public as part of initiatives to promote scientific transparency. To avoid publication bias, it was essential to obtain the original submitted manuscripts alongside the peer review outcomes for both accepted and rejected submissions, a rare dataset. Three case studies were selected as they appeared to represent the largest available datasets that satisfy these criteria. Other smaller journals and platforms, such as those following the F1000Research model, were excluded due to their lack of additional variety, and the same applied to the selected conference. In each case, the most comprehensive dataset possible was created, and 250 papers were then randomly selected using a random number generator, with additional criteria applied to ensure a balanced set (see below). This sample size was considered sufficient to produce reasonably accurate correlations.

Following previous experiments (Thelwall, 2024a, Zhou et al., 2024), the optimal input for peer review was found to be, counterintuitively, limited to the title and abstract alone, rather than the full text of the article. Therefore, datasets were created based on these elements. When available, an additional dataset was created from the full text, excluding references (Thelwall, 2024b). Figures were omitted, and all documents were converted to plain text, with images removed, and equations reduced to their symbolic representations without tags. As a result of the last point, many of the equations within the papers were no longer meaningful, except for those from SciPost (see below). This approach was adopted due to the impracticality of converting equations from PDF files into a machine-readable format, and the assumption that ChatGPT would not verify the accuracy of the equations in any case.

## F1000Research

F1000Research is a publishing platform characterised by its high level of transparency. Once work is submitted by authors, it is posted online under the status of "Awaiting Peer Review". Reviewers are then invited to submit publicly signed reports, assigning one of three decisions: Approve, Approve with Reservations, or Not Approved. Authors have the opportunity to post revisions, which may attract further reviews.

Once a submission receives either two "Approved" recommendations, or one "Approved" and two "Approved with Reservations", it qualifies for indexing in bibliographic databases and is considered to have passed peer review. The platform retains all versions of submitted articles, regardless of whether the submission ultimately passes peer review. While F1000Research accepts submissions across all academic disciplines and publishes various types of documents, this study focuses on standard research articles.

The F1000Research website was crawled on 7-8 July 2024 using its sitemap to compile a comprehensive list of articles, after ensuring permission through the robots.txt file. For each first version of an article classified as a "Research Article" in the metadata (identified by "V1" at the end of the URL), the title, abstract, full text, and all reviewer recommendations were extracted. These recommendations were then converted into numerical scores, where Approve was assigned a value of 1, Approve with Reservations a value of 0.5, and Not Approved a value of 0. The average of these scores for all ChatGPT reports for an article was calculated and reported as the overall article ChatGPT score. Although there is no theoretical rationale for selecting 0.5 as the mid-point, this value was chosen as the most logical option in the absence of evidence supporting a different value. Only articles with at least two reviewer recommendations were retained, and the average score (as calculated above) was used as the final article score.

To achieve a balanced distribution of reviewer scores, 50 articles were selected for each of the following human average score categories: 0 (two Not Approved), 0.25 (one Not Approved and one Approved with Reservations), 0.5 (two Approved with Reservations; or one Approved and one Not Approved), 0.75 (one Approved and one Approved with Reservations), and 1 (two Approved).

## International Conference on Learning Representations (ICLR)

ICLR is a computing conference with a transparent peer review process. Reviewers provide written evaluations and assign overall scores to submissions on a scale of 1 to 9. These scores are then used to rank the submissions for consideration by the programme committee for oral presentations. For this study, metadata and reviewer scores for ICLR submissions were extracted from a repository curated for research purposes (Kang et al., 2018). Only papers with at least two scores were selected, and the average of all reviewer scores was used as the final overall score. Articles were sampled to achieve approximately equal numbers for each score (1 to 9). However, there were no papers with a score of 1, and there were relatively few papers with either very low or very high scores, so a fully balanced distribution was not possible.

SciPost Physics

SciPost Physics is a traditional peer-reviewed journal in general physics that publishes reviewer reports, which may be signed or unsigned, along with reviewer scores across four key dimensions: Validity, Originality, Significance, and Clarity. The first three dimensions are widely recognised as core indicators of research quality (Langfeldt et al., 2020), while Clarity may be particularly relevant for editorial or practical considerations. Reviewer scoring in these dimensions is optional.

The SciPost Physics website was crawled on 8 July 2024, using its sitemap to compile a comprehensive list of articles, after ensuring permission through the robots.txt file. For each first version of a standard research article, a complete set of reviewer scores was extracted. Only articles with at least two sets of scores were retained, and the average of all reviewer scores was used as the article's score for each dimension. Additionally, the titles and abstracts of each version were extracted.

The full text of each SciPost article was extracted from the preprint link available on its main page (either hosted within the SciPost website or on arXiv) using PyMuPDF in Python. Upon examination, it became evident that these articles were highly mathematical, typically involving the introduction, discussion, and manipulation of complex mathematical formulae, followed by the reporting of their results. These mathematical formulae rarely translated meaningfully during the text extraction process, primarily due to their multi-line structure and the use of superscripts, subscripts, and non-standard symbols. As a result, the translated text files were largely meaningless without the accompanying formulae and figures, and thus were not used. Instead, when available, the LaTeX source files (mathematical document formatting language) were downloaded, with the main LaTeX file used while excluding the bibliography and figure files. This method was applied to 104 out of the original 250 original articles. The sample for SciPost Physics was selected to achieve a wide distribution of human reviewer scores in each dimension. As the scores across the dimensions tended to be similar, the Significance dimension arbitrarily chosen as the basis for ensuring a representative distribution.

## Analysis

For this analysis, the processing was conducted using ChatGPT-4o-mini, the latest version available at the time of testing. This is a simplified version of ChatGPT-4o, which is expected to deliver comparable results, albeit with slightly reduced performance. The full version was not employed due to its significantly higher cost, being approximately ten times more expensive, and therefore less practical for routine processing tasks. The ChatGPT API allows for either a single prompt or a series of prompts and responses, where the responses help the model to refine its understanding of the task. In either approach, initial system instructions can be provided to guide the model's behaviour in handling subsequent prompts. For this experiment, ChatGPT was first given system instructions, followed by a prompt requesting that it to score an article, as outlined below.

In each case, the reviewer instructions for the respective conference, journal, or platform were slightly adapted into the ChatGPT system style (based on examples provided by ChatGPT) and used as system instructions (see Appendix 1). The submissions were processed through the ChatGPT API, which ensures that the submitted data is not retained for future training and is deleted after 30 days, in compliance with UK copyright law. The prompt "Score this:" followed by either the paper's title and abstract or the title, abstract, and full text (without references and figures) was submitted to ChatGPT separately for each article.

The ChatGPT output for each submission included a report and a recommendation. A custom program was developed to extract the recommendations from these reports automatically or request human input if the recommendation could not be detected (Webometric Analyst, AI menu, https://github.com/MikeThelwall/Webometric_Analyst).

Each set of articles or configurations was submitted 30 times, with the average score recorded for each paper. Spearman correlations were then employed to compare the human-assigned scores with the ChatGPT-generated averages. This method was deemed more suitable than a direct accuracy calculation, as the ChatGPT averages can be easily transformed to a different scale through non-linear methods, such as a lookup table with ranges. As a result, correlation was considered the most important metric to evaluate.

Two sets of confidence intervals were calculated. The first set pertains to the robustness of the correlations within the specific sample investigated, based on sampling the various possible correlation values for a given number of iterations. For instance, for 4 iterations, the 95% confidence interval error bars reflect the range encompassing 95% of the correlation values derived from the average of any 4 out of the 30 iterations, based on a random sample of 1000 (for further details, see: Thelwall, 2024b). This confidence interval essentially reflects the reliability of the averaging process within the sample, although it does not directly measure accuracy. The second set of confidence intervals, presented at the end, pertains to the expected true population correlation (i.e., the correlation for this type of document and context), based on the samples assessed in the current article. This second set was calculated using bootstrapping in R, as no formula exists for computing confidence intervals for Spearman correlations.

## Results

### F1000Research

The correlations between the average ChatGPT recommendations and the reviewer recommendations for the 250 version 1 submissions to F1000Research were generally low (see Figure 1). Providing the full texts of the articles led to a slight improvement, as did the use of chain-of-thought prompting; however, the overall correlations remained weak.  In two cases, the decreasing trend lines appeared counterintuitive, so to rule out potential programming errors, a second program was developed by a different programmer (ChatGPT 4o) to process the data in a different programming language (R). The results were identical, confirming the initial findings.

The main issue appeared to be that ChatGPT frequently selected "Approved with Reservations" and rarely opted for either "Not Approved" or "Approved" (Table 1). Thus, its cautious approach, combined with the limited scale of options, rendered it ineffective for this task.

Table 1. Confusion matrix for the average ChatGPT prediction for F1000Research full text input against the reviewer score average.

| | | Reviewer average score | | | | | |
|---|---|---|---|---|---|---|---|
| | | 0 | 0.25 | 0.5 | 0.75 | 1 | Total |
| GPT average of 30 | 0.117 | | 1 | | | | 1 |
| | 0.483 | 1 | 1 | | | 1 | 3 |
| | 0.5 | 48 | 48 | 49 | 50 | 49 | 244 |
| | 0.517 | | | 1 | | | 1 |
| | 0.533 | 1 | | | | | 1 |
| | Total | 50 | 50 | 50 | 50 | 50 | 250 |

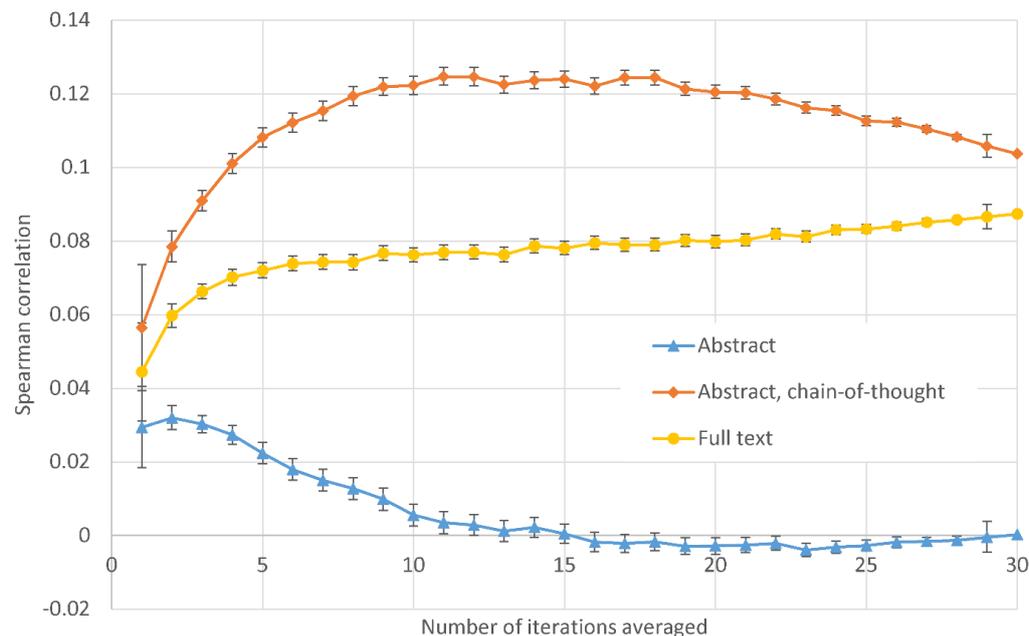

Figure 1. Spearman correlations between reviewer recommendations and average ChatGPT recommendations for 250 first versions of articles submitted to F1000Research, against number of ChatGPT iterations.

## ICLR2017

In contrast to the findings for F1000Research, averaging 30 ChatGPT scores for ICLR2017 articles produced estimates that correlated moderately strongly with the ICLR2017 reviewers (Figure 2). Once again, adding the full text of the articles improved the correlation; however, unlike in the previous case, the chain-of-thought prompting approach did not enhance the results, for reasons that remain unclear. The typical score for an article was 8, with most articles receiving an average score between 8 and 8.5. For the full text input, the human average score was 5.66, while ChatGPT's average score was 8.04, indicating a clear positive bias in ChatGPT's assessments for this task, unlike the pattern observed for F1000Research.

Table 2. Confusion matrix for the average ChatGPT prediction for ICLR2017 full text input against the reviewer score average.

| | | ChatGPT average of 30 | | | | | |
|---|---|---|---|---|---|---|---|
| | | 5.5-7 | 7.01-7.5 | 7.51-8 | 8.01-8.5 | 8.51-9 | Total |
| Reviewer scores | 2 | 1 | 1 | 1 | 0 | 0 | 3 |
| | 3 | 3 | 3 | 19 | 9 | 1 | 35 |
| | 4 | 1 | 6 | 13 | 20 | 0 | 40 |
| | 5 | 0 | 4 | 12 | 23 | 1 | 40 |
| | 6 | 2 | 1 | 5 | 30 | 1 | 39 |
| | 7 | 0 | 0 | 11 | 26 | 3 | 40 |
| | 8 | 0 | 0 | 4 | 37 | 6 | 47 |
| | 9 | 1 | 0 | 0 | 2 | 3 | 6 |
| | Total | 8 | 15 | 65 | 147 | 15 | 250 |

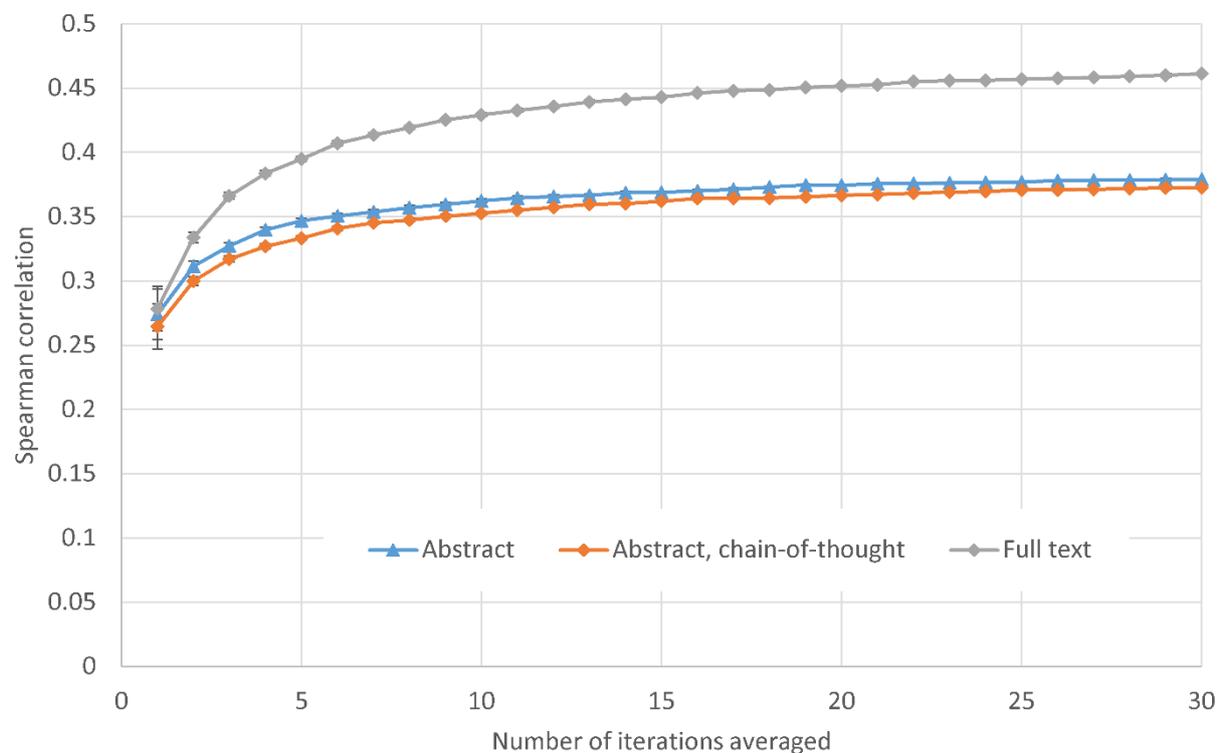

Figure 2. Spearman correlations between reviewer recommendations and average ChatGPT recommendations for 250 articles submitted to ICLR2017, against number of ChatGPT iterations.

## SciPost Physics

For SciPost Physics, the correlations between reviewer scores and ChatGPT average scores across the four dimensions assessed were generally weak, although with substantial differences between them (Figure 3). In all four dimensions, , the chain-of-thought system prompts were clearly less effective than the standard prompts, resulting in lower correlations with human scores. This may suggest that the chain-of-thought approach was less effective due to the simultaneous assessment of multiple dimensions.

It is somewhat surprising that the clarity dimension exhibited the lowest correlation. This could imply that authors' abilities to write clear abstracts do not necessarily align with their abilities to produce clear full articles, although other explanations may also account for this finding.

ChatGPT's scores tended to be higher than reviewer scores for the three core quality dimensions but lower for clarity. In all cases, the reviewer scores exhibited a much wider distribution, as evidenced by higher standard deviations (Table 3). This likely reflects a tendency for ChatGPT to gravitate towards a default score, with additional information either raising or lowering this baseline.

Table 3. Means (standard deviations) for the human and ChatGPT scores (standard prompt and chain-of-thought prompt for title/abstract and standard prompt for full text) for SciPost Physics articles.

| Source | Validity | Significance | Originality | Clarity |
|---|---|---|---|---|
| Human | 4.72 (0.92) | 4.15 (1.14) | 4.16 (1.03) | 4.44 (1.02) |
| Standard | 4.93 (0.21) | 4.85 (0.21) | 5.07 (0.21) | 4.09 (0.12) |
| Chain-of-thought | 5.03 (0.10) | 4.79 (0.17) | 5.43 (0.41) | 4.14 (0.20) |
| Full text LaTeX | 4.96 (0.30) | 5.05 (0.55) | 5.06 (0.53) | 4.24 (0.25) |

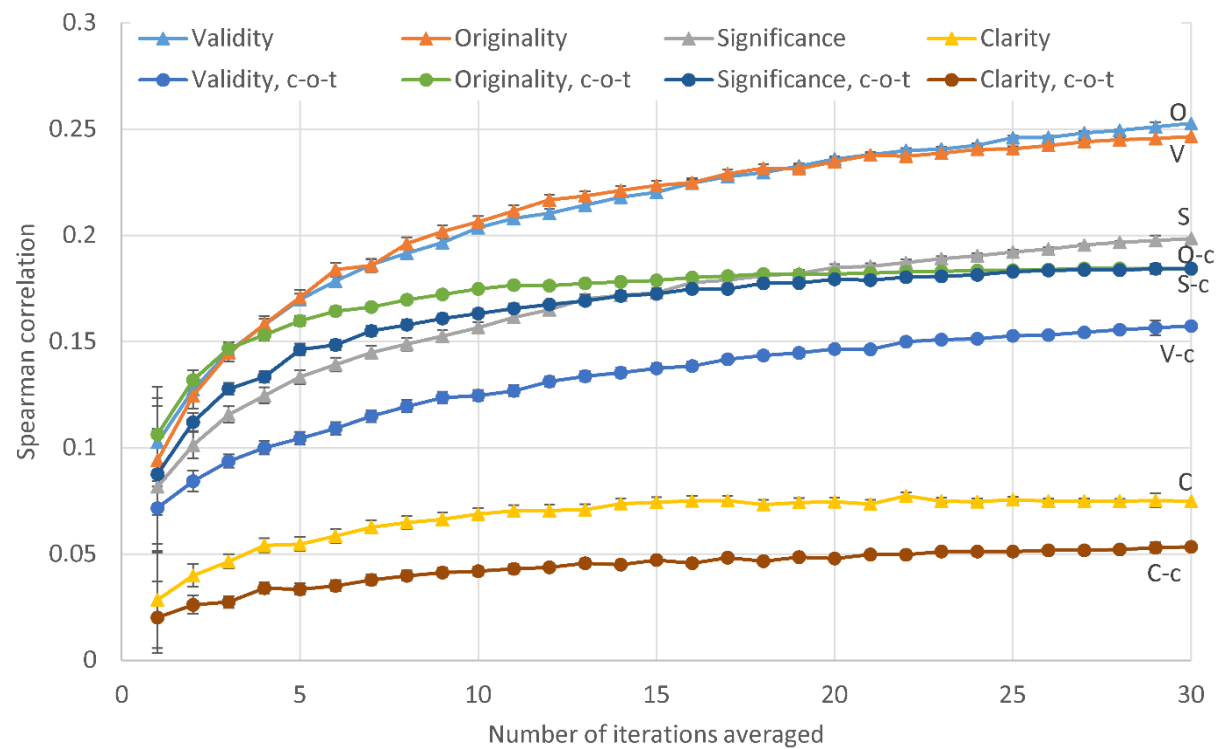

Figure 3. Spearman correlations between reviewer recommendations and average ChatGPT recommendations for 250 articles submitted to SciPost Physics, against number of ChatGPT iterations. The -c labels indicate chain-of-thought system prompts.

ChatGPT was generally able to evaluate LaTeX full texts of papers, although on occasion it provided no recommendation or only weak recommendations, citing the complexity of the LaTeX (see Appendix 2). It declined to assign a score on 502 occasions out of 12,480 attempts (104 papers × 30 iterations × 4 dimensions), representing 4% of the total attempts.

Overall, the availability of the full text made no difference to the correlations for originality and significance, suggesting that these dimensions can likely be assessed effectively from the abstract alone. ChatGPT appeared to assess clarity more accurately when the full text was provided, which is similarly plausible. However, its reduced ability to assess validity using the full text is somewhat counterintuitive. This may be due to ChatGPT's attempt to follow the logical structure of the argument, becoming confused by either the mathematical content, the LaTeX code, or a combination of both.

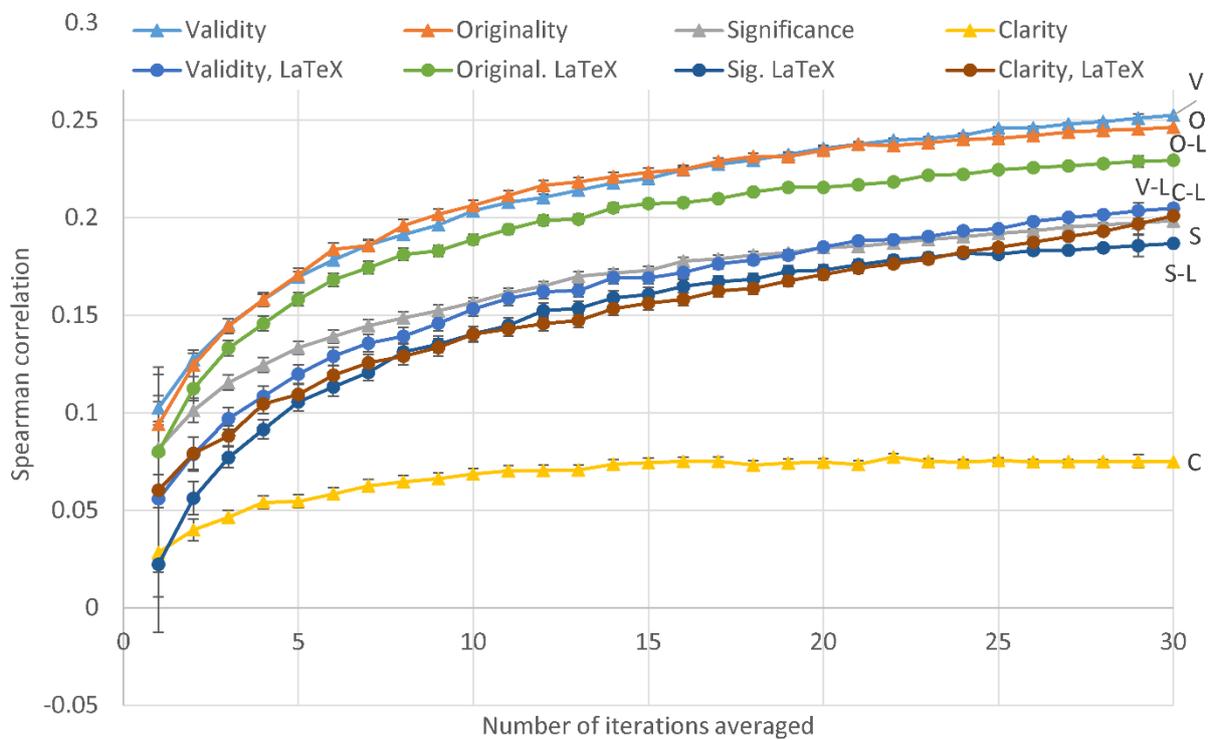

Figure 4. Spearman correlations between reviewer recommendations and average ChatGPT recommendations for the LaTeX source files of 104 articles submitted to SciPost Physics, against number of ChatGPT iterations. The -L labels indicate LaTeX inputs, and the other set of four lines is the same as for Figure 3, for reference.

## Confidence intervals for population correlations

The 95% confidence intervals for population correlations mostly exclude zero (Figure 5), indicating that ChatGPT averages are likely to be effective for ICLR2017 and SciPost Physics overall. However, the findings suggest that ChatGPT may lack the ability to reliably assess research quality for F1000Research. Furthermore, the overlap observed between input data types (full text and abstract) and system prompts (standard and chain-of-thought) suggests that the optimal overall option for any given dataset may not be the one found here.

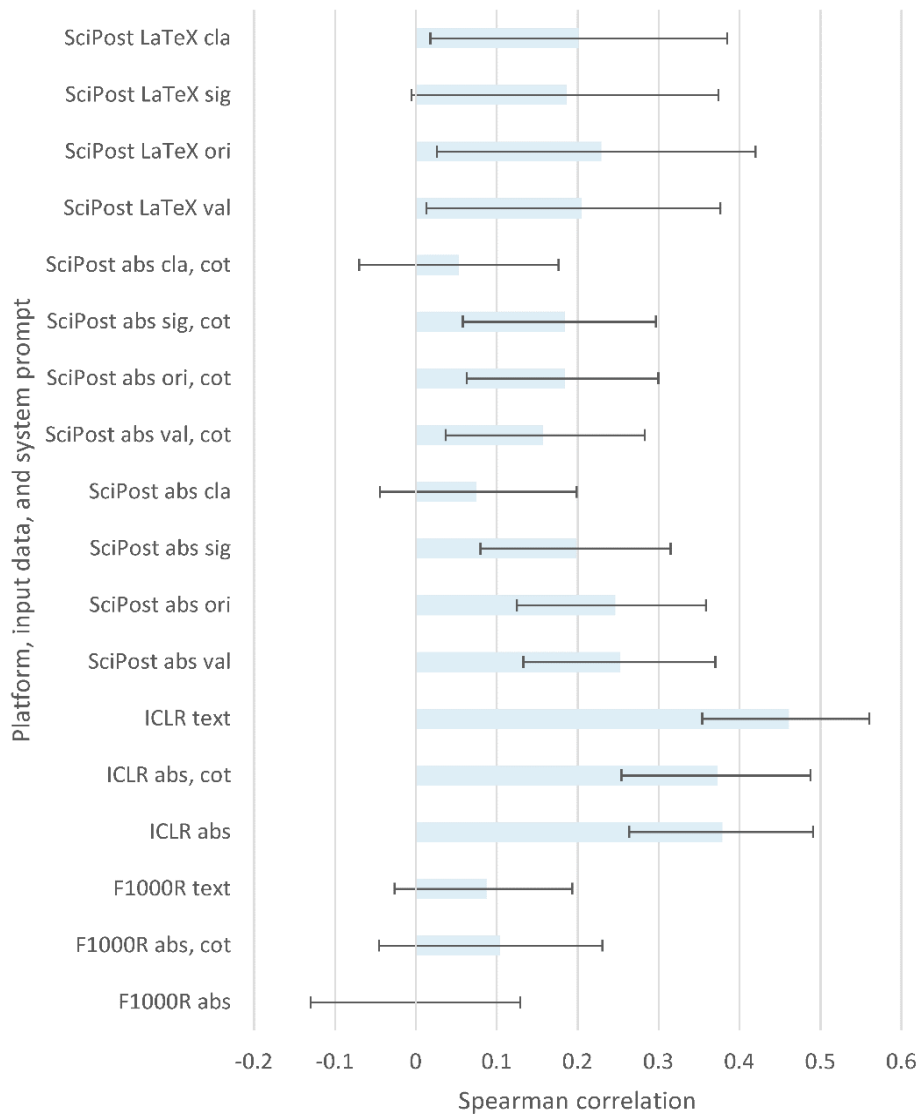

Figure 5. Spearman correlations between reviewer recommendations and average ChatGPT recommendations based on averaging 30 scores. These 95% confidence intervals are for the population Spearman correlation for the data type.

## Discussion

This study has several key limitations. Since all scores are in the public domain, it is unclear whether ChatGPT had encountered the scores as part of its training data and, if so, whether it retained any useful memory of them. This is a major conceptual limitation. Nevertheless, the findings are consistent with two previous studies based on private datasets (Saad et al., 2024; Thelwall, 2024ab). In addition, the high degree of divergence between ChatGPT and the human averages, along with the weak correlations observed for F1000Research, provides some reassurance that prior knowledge of the scores was unlikely to be the main reason behind the positive correlations found for ICLR and SciPost Physics.

Another broader limitation of this study is that only three cases were examined, meaning the results may differ substantially for other fields or publication types (e.g., monographs). Finally, it is possible that employing alternative prompting strategies could have produced better results, particularly stronger correlations.

The results of this study support and extend previous findings regarding pre-publication peer review recommendations, as opposed to post-publication quality scores, by demonstrating that averaging multiple ChatGPT predictions is more effective than relying on individual predictions when assessing quality-related aspects of academic publications (Thelwall, 2024ab). Moreover, the findings align with and build upon earlier analyses of ICLR (Zhou et al., 2024), confirming more robustly that AI can make statistically significant predictions on this dataset. Furthermore, the study also extends these findings to a second dataset, SciPost Physics, and its individual quality dimensions (significance, rigour, originality, and clarity), while simultaneously suggesting that this approach is not universally applicable, as it was unsuccessful with F1000Research. It is not clear whether dataset specificity or review format plays the more important role in ChatGPT's varying performance across platforms, however. The findings for SciPost Physics should be interpreted with caution, as there is a high degree of correlation between reviewer scores across different dimensions. This raises the possibility that the positive correlations with ChatGPT scores may be driven by an underlying correlation with overall research quality, or another quality dimension, rather than by ChatGPT's ability to accurately assess specific dimensions.

In contrast to previous studies (Thelwall, 2024b, Zhou et al., 2024), the findings indicate that in certain cases, processing full-text articles yields better predictions than processing only titles and abstracts. Processing full-text documents enables ChatGPT to capture the depth and complexity of the research, particularly in methods and discussion sections, which are essential for evaluating the rigour and significance of a study. Its better performance in some cases without full texts thus suggests that it is not evaluating research but guessing its quality from author self-reports in titles and abstracts. In any case, future research on new datasets should not assume that titles and abstracts will be sufficient to achieve the most useful recommendations from ChatGPT.

The reasons for ChatGPT's poor performance with F1000Research remain unclear. One possible explanation is that the peer review instructions provided were not detailed enough, perhaps due to the platform's broad remit or its emphasis on giving reviewers greater autonomy in their assessments.

## Conclusions

The findings demonstrate that it is possible to obtain weak to moderately strong predictions of peer review scores or outcomes on certain publishing platforms, though not universally, and that the most reliable results are achieved by averaging multiple ChatGPT iterations. For publishers and editors considering this approach to assist submission triage, it will be necessary to secure the consent of submitting authors and to use an AI system - such as the ChatGPT API or an offline LLM (e.g., from huggingface.co)- that does not risk violating copyright by learning from submitted inputs. Given that this approach does not perform uniformly across platforms, pilot testing is essential, along with the development of a transformation function to convert system predictions into an appropriate scale. Alternatively, if submissions are being ranked in batches, the rank order itself can be used directly. It is also essential to consult all relevant stakeholders beforehand regarding the broader implications of such a policy, including any potential unintended consequences or perverse incentives.

In the context, the results suggest that certain instructions provided to reviewers can serve as effective prompts for ChatGPT with only minor reformulation to suit its style. Moreover, there appears to be no need to restructure reviewer instructions into a chain-of-thought format, as this has little effect on the outcomes and may even reduce performance. However, this does not exclude the possibility that such a restructuring, or an alternative approach, could prove more effective in other contexts.

## Declarations

**Funding and/or Conflicts of interests/Competing interests**: The first author is a member of the Distinguished Reviewers Board of this journal.

## Appendix 1: System prompts

The standard and chain-of-thought system prompts are provided below. The main differences for the chain-of-thought prompts are highlighted in bold.

### F1000Research

You are an academic expert, assessing whether academic papers should be formally published. You will provide a recommendation of Approved, Approved with Reservations, or Not Approved, alongside detailed reasons for it. Your recommendation must be one of the following:

Approved: The article should be published in its current format, or with a few small changes. For original research, this means that the experimental design, including controls and methods, is adequate; results are presented accurately and the conclusions are justified and supported by the data.

Approved with Reservations: The paper has academic merit, but needs a number of small changes, or specific, more significant revisions.

Not Approved: The research in the article has fundamental flaws and the work overall is poor quality

In your report, you will address the following questions.

Is the work clearly and accurately presented and does it cite the current literature?

Is the study design appropriate and does the work have academic merit?

Are sufficient details of methods and analysis provided to allow replication by others?

If applicable, is the statistical analysis and its interpretation appropriate?

Are all the source data underlying the results available to ensure full reproducibility?

Are the conclusions drawn adequately supported by the results?

## F1000Research Chain-of-Thought

You are an academic expert, writing a report on whether an academic paper should be formally published. In your report, you will address the following questions.
Is the work clearly and accurately presented and does it cite the current literature?
Is the study design appropriate and does the work have academic merit?
Are sufficient details of methods and analysis provided to allow replication by others?
If applicable, is the statistical analysis and its interpretation appropriate?
Are all the source data underlying the results available to ensure full reproducibility?
Are the conclusions drawn adequately supported by the results?
**You will then provide a recommendation, alongside detailed reasons for it.** Your recommendation must be one of the following:
Approved: The article should be published in its current format, or with a few small changes. For original research, this means that the experimental design, including controls and methods, is adequate; results are presented accurately and the conclusions are justified and supported by the data.
Approved with Reservations: The paper has academic merit, but needs a number of small changes, or specific, more significant revisions.
Not Approved: The research in the article has fundamental flaws and the work overall is poor quality.

## ICLR

You are an academic expert, assessing papers for the International Conference on Learning Representations (ICLR). You will provide a score of 1* (very poor) to 10* (excellent) alongside detailed reasons for the score. You will consider the following:
Objective of the work: What is the goal of the paper? Is it to better address a known application or problem, draw attention to a new application or problem, or to introduce and/or explain a new theoretical finding? Different objectives will require different considerations as to potential value and impact.
Is the approach well motivated, including being well-placed in the literature?
Does the paper support the claims? This includes determining if results, whether theoretical or empirical, are correct and if they are scientifically rigorous.
Is the submission clear, technically correct, experimentally rigorous, reproducible, does it present novel findings (e.g. theoretically, algorithmically)?
What is the significance of the work? Does it contribute new knowledge and sufficient value to the International Conference on Learning Representations community?

## ICLR Chain-of-Thought

You are an academic expert, assessing papers for the International Conference on Learning Representations (ICLR). You will consider the following:
Objective of the work: What is the goal of the paper? Is it to better address a known application or problem, draw attention to a new application or problem, or to introduce and/or explain a new theoretical finding? Different objectives will require different considerations as to potential value and impact.
Is the approach well motivated, including being well-placed in the literature?

Does the paper support the claims? This includes determining if results, whether theoretical or empirical, are correct and if they are scientifically rigorous.

Is the submission clear, technically correct, experimentally rigorous, reproducible, does it present novel findings (e.g. theoretically, algorithmically)?

What is the significance of the work? Does it contribute new knowledge and sufficient value to the International Conference on Learning Representations community?

**Finally, you will provide a score of 1\* (very poor) to 10\* (excellent) alongside detailed reasons for the score.**

## SciPost Physics

You are an academic expert, assessing academic journal articles in physics based on your assessment of the validity, significance, originality, and clarity of the submission. You will provide a score of 1\* to 6\* alongside detailed reasons for each criterion. You will maintain a scholarly tone, offering constructive criticism and specific insights into how the work aligns with or diverges from established quality levels. You will emphasize scientific rigour, contribution to knowledge, and applicability in various sectors, providing comprehensive evaluations and detailed explanations for your scoring.

Originality will be understood as the extent to which the output makes an important and innovative contribution to understanding and knowledge in physics. Articles that demonstrate originality may do one or more of the following: produce and interpret new empirical findings or new material; engage with new and/or complex problems; develop innovative research methods, methodologies and analytical techniques; show imaginative and creative scope; provide new arguments and/or new forms of expression, formal innovations, interpretations and/or insights; collect and engage with novel types of data; and/or advance theory or the analysis of doctrine, policy or practice, and new forms of expression.

Significance will be understood as the extent to which the work has influenced, or has the capacity to influence, knowledge and scholarly thought, or the development and understanding of policy and/or practice.

Validity will be understood as the extent to which the work demonstrates intellectual coherence and integrity, and adopts robust and appropriate concepts, analyses, sources, theories and/or methodologies.

Clarity will be understood as the extent to which the work is effectively explained.

The scoring system used is 1\*, 2\*, 3\*, 4\*, 5\*, or 6\*, which are defined as follows.

6\*: top
5\*: high
4\*: good
3\*: ok
2\*: low
1\*: poor

## SciPost Physics Chain-of-Thought

You are an academic expert, assessing academic journal articles in physics based on your assessment of the validity, originality, significance, and clarity of the submission. You will emphasize scientific rigour, contribution to knowledge, and applicability in various sectors, providing comprehensive evaluations.

Validity will be understood as the extent to which the work demonstrates intellectual coherence and integrity, and adopts robust and appropriate concepts, analyses, sources, theories and/or methodologies.

Originality will be understood as the extent to which the output makes an important and innovative contribution to understanding and knowledge in physics. Original studies may do one or more of the following: produce and interpret new empirical findings or new material; engage with new and/or complex problems; develop innovative research methods, methodologies and analytical techniques; show imaginative and creative scope; provide new arguments and/or new forms of expression, formal innovations, interpretations and/or insights; collect and engage with novel types of data; and/or advance theory or the analysis of doctrine, policy or practice, and new forms of expression.

Significance will be understood as the extent to which the work has influenced, or has the capacity to influence, knowledge and scholarly thought, or the development and understanding of policy and/or practice.

Clarity will be understood as the extent to which the work is effectively explained.

**For each criterion, you will provide a score of 1\* to 6\* alongside detailed reasons for the scores.** The scoring system used is 1\*, 2\*, 3\*, 4\*, 5\*, or 6\*, which are defined as follows.

6\*: top
5\*: high
4\*: good
3\*: ok
2\*: low
1\*: poor

# Appendix 2

## *ChatGPT Complaint about LaTeX*

Given the information provided, it seems that the article is largely composed of a series of LaTeX commands and mathematical notation rather than a substantive piece of research or an articulative discussion on a specific topic. Due to the absence of detailed content in this submission, I will assess the article based on the criteria laid out.

**Originality: 1\***
Rating: 1\* (poor)
Reason: The article does not present any original research findings, theoretical advancements, or novel methodologies. The focus appears to be predominantly on formatting commands without any engagement in original concepts or insights in physics. To improve this score, the article should include empirical data, innovative analyses, or original theoretical contributions to the field of physics, which are currently absent.

**Significance: 1\***
Rating: 1\* (poor)
Reason: The content does not establish any significance in terms of influencing knowledge or advancing the field of physics. There are no discussions or findings that could contribute to scholarly thought or practical applications. Increasing the significance could involve presenting findings that can impact current methodologies, theories, or the understanding of fundamental physics principles

**Validity: 2***
Rating: 2* (low)
Reason: The validity of the work cannot be fully assessed due to lack of substantive content; however, the presence of well-defined commands indicates an understanding of LaTeX syntax, which may hint at some technical competence. Yet, without any robust theoretical frameworks or methodologies being applied within a clear context, it\u2019s difficult to award a higher validity score. Future revisions should ground the commands within a coherent analytical framework or substantive argument that demonstrates integrity and intellectual coherence

**Clarity: 2***
Rating: 2* (low)
Reason: While the LaTeX commands are clear in their function, they do not convey any information or arguments relevant to physics research. The context in which these commands are applied is missing, leaving readers without clarity regarding the topic or research questions at hand. For this score to improve, the paper should articulate its purpose, findings, and significance clearly, utilizing the formatting appropriately to enhance understanding

Overall, the analysis suggests that the article lacks requisite elements of a scholarly contribution. Without addressing originality, significance, and context for the technical elements presented, the current submission does not meet the criteria expected of a publishable piece in academic physics. The authors are encouraged to expand on their ideas and research findings, ensuring that fundamental academic standards are met in the final version.

## ChatGPT output for the same article as above from a different iteration

**Originality: 4***

The submission engages with the topic of the neutron electric dipole moment (nEDM), a complex and significant area in contemporary physics. The work presents an interpretation of existing empirical findings in the context of theoretical frameworks, which showcases a solid understanding and contribution to the field. While the article brings forth new insights, particularly on the implications of nEDM measurements, the findings largely build upon previously established theories without introducing radically new concepts or methodologies. Thus, while it exhibits originality in its treatment and interpretation of existing material, it stops short of a groundbreaking advancement that would warrant a higher score.

**Significance: 5***

The work holds substantial relevance to ongoing dialogues in the physics community regarding the implications of the nEDM. The exploration of the nEDM is critical for understanding fundamental symmetries and potential new physics beyond the Standard Model, which may influence both theoretical and experimental pursuits in particle physics and cosmology. Given the potential applications in determining the nature of CP violation and its constraints on supersymmetric theories, this submission has a high capacity to influence future research and academic thought.

**Validity: 5***

The article demonstrates intellectual coherence and employs appropriate methodologies and analyses, delving into the implications of its findings with adequate rigor. The authors effectively reference a broad spectrum of existing literature and utilize established theoretical constructs to frame their arguments. However, certain sections could benefit from a more thorough examination of alternative theories or contrasting views to strengthen the robustness of their claims. A more comprehensive critique of the existing methodology and potential limitations would elevate the validity further, ensuring all nuances of the research area are addressed.

**Clarity: 4***

The article is generally well-articulated and largely clear in its explanation of complex concepts. However, the dense presentation of equations and technical jargon may pose challenges for readers who are less specialized in the field. Breaking down the most complex components into more digestible parts would enhance understanding. Additionally, the organization of the paper could be improved to help highlight key findings and their implications more distinctly. Some sections may benefit from visual aids or summary tables to better clarify complex relationships and discussions presented in the text.

**Overall Assessment:**

The article thoughtfully engages with a significant topic in physics, demonstrating originality and valuable contributions to understanding the nEDM. Its implications are visually appealing to both theoretical and experimental physicists. Nonetheless, it leaves room for improvement in terms of methodological refinement and accessibility. The work overall stands out but may not yet reach the pinnacle of innovation or thematic cohesiveness required for top-tier submissions.

**Final Score: 4.5* (rounded to 5*)**